# Dual-Polarization Phase Retrieval Receiver in Silicon Photonics

Brian Stern, Hanzi Huang, Haoshuo Chen, Kwangwoong Kim, and Mohamad Hossein Idjadi

*Nokia Bell Labs, Murray Hill, New Jersey, USA*

**Abstract**

We demonstrate a silicon photonic dual-polarization phase retrieval receiver. The receiver recovers phase from intensity-only measurements without a local oscillator or transmitted carrier. We design silicon waveguides providing long delays and microring resonators with large dispersion to enable symbol-to-symbol interference and dispersive projection in the phase retrieval algorithm. We retrieve the full field of a polarization-division multiplexed 30-GBd QPSK and 20-GBd 8QAM signals over 80 km of SSMF.

## Introduction

Short-reach optical communication continues to push to higher capacities while also demanding low-cost devices [1,2]. Coherent optics provide benefits such as receiver sensitivity and spectral efficiency to support high capacity, but the requirement of a wavelength-aligned local oscillator (LO) at the receiver adds complexity and cost to these devices. On the other hand, direct-detection systems, which only measure the intensity at the receiver, are simple and low cost, but they face capacity limitations, being unable to detect complex-valued modulation formats and also facing spectral fading due to chromatic dispersion in fiber [3]. There are also self-coherent schemes which transmit a carrier with the signal instead of using an LO. However, the carrier consumes transmitter signal power, and a high carrier-signal power ratio (CSPR) can lead to nonlinear impairments during fiber transmission and reduced receiver sensitivity [4]. Similarly, single-sideband detection systems also require co-transmitting a carrier [5,6].

The recently proposed phase retrieval receiver upends many of these difficult trade-offs [7–

9]. This receiver can recover the phase of an incoming signal, including one encoded in complex-valued modulation formats, without an LO or transmitted carrier of any kind, despite using direct detection. A phase retrieval receiver uses a combination of dispersive propagation and symbol-to-symbol interference to reconstruct the full field, including phase, from the intensity-only measurements [10]. The wavelength alignment is relaxed compared to coherent, and requiring only a passive receiver can support lower-cost devices. Phase retrieval is robust in the presence of chromatic dispersion in fiber, and finally, because no carrier is needed, the CSPR is not a constraint.

Silicon photonics is an ideal platform on which to implement these receiver technologies in massively manufacturable, low-cost devices [11]. Coherent and self-coherent receivers requiring an LO or carrier have previously been demonstrated using silicon photonics [12–19]. We recently reported a carrier-less silicon photonic phase retrieval receiver for a single polarization [20]. However, practical systems require handling changes in polarization during transmission. Further, leveraging both polarizations doubles the total capacity.

Here we report the first fully integrated, carrier-less silicon photonic phase retrieval receiver using polarization-division multiplexing (PDM). The chip includes silicon microring resonators, waveguide delay lines, and integrated high-speed photodiodes. We show transmission of dual polarization (DP) quadrature phase shift keying (QPSK) signals and DP-8QAM (quadrature amplitude modulation) signals over 80 km of standard single mode fiber (SSMF) using the integrated receiver.

## Phase Retrieval with Direct Detection

In order to retrieve the full fields of signals using direct detection without a carrier, we employ a space-time diversity phase retrieval approach [8]. In summary, the two polarizations are first demultiplexed using a polarization beam splitter rotator (PBSR) (Fig. 1). Next, each polarization is processed as four intensity-only measurements. One of these four branches includes a large dispersion *D* to relate the phase to an amplitude modulation. The other three branches include varying time delays $\Delta t_i$ to produce symbol-to-symbol interference. Using these measurements, the digital signal processing (DSP) uses a modified Gerchberg-Saxton (GS) algorithm to retrieve the signals' full fields [8].

## Polarization, Dispersion, and Delays On-Chip

We implement the phase retrieval receiver in silicon by designing a photonic circuit with the three key components of the approach: polarization management, large dispersion, and long delay lines.

The input to the chip is an edge-coupled inverse taper waveguide which provides low-loss coupling for both polarizations (Fig. 2). Next, a PBSR splits the two polarizations and converts both to the quasi-transverse electric (TE) mode. The loss of the silicon PBSR is lower than 1 dB and the polarization crosstalk is lower than -20 dB. Then, each channel is sent to a separate, mirrored copy of the subsequent elements of the receiver.

We achieve a large on-chip dispersion by designing microring resonators to introduce spectrally varying group delay [20,21]. Figure 3(a) shows our approach. The three silicon resonators are all strongly over-coupled ($\kappa_1$=23%, $\kappa_2$=31%, and $\kappa_3$=39%) and their resonances are detuned from each other. This results in a large peak group delay for each resonance with minimal extinction ratio (Fig. 3(b)). The simulated insertion loss is around 0.2 dB and the dispersion (slope of the group delay) is -95 ps/nm, which is sufficient for space-time diversity phase retrieval. Integrated heaters are used to tune the rings, but owing to their intentionally low *Q*, the performance is not overly sensitive to alignment. Large on-chip dispersion could also be achieved using chirped Bragg gratings [22].

The delay lines must incorporate long waveguide lengths with very low loss in a small area on-chip. To accomplish this, we use wide multimode waveguides to minimize propagation loss from the sidewalls. Additionally, to avoid exciting higher-order modes, we use Euler bends rather than circular arcs. By spiraling the waveguides into a compact shape, we fit delay lengths of $L_2 = 1.3$ cm and $L_3 = 2.2$ cm, or ~120 ps and ~200 ps, respectively (Fig. 2). These correspond to symbol-to-symbol interference of 3.5 and 6 symbol periods at 30 GBd.

The remainder of the receiver in Fig. 2 consists of splitters of photodiodes. Each polarization signal is split with a 1x4 multimode interferometer (MMI) to the dispersion and time delay branches. The three delay line branches interfere using a 3x3 MMI. The MMIs have simulated losses of 0.14 and 0.13 dB, respectively. Each branch is detected by an integrated germanium photodiode with a nominal bandwidth over 40 GHz.

## Fabrication and Assembly

The phase retrieval receiver is fabricated at a foundry on 300 mm silicon-on-insulator (SOI) wafers with 220 nm device layer thickness.

We assemble the fabricated device (3.3 mm × 1.4 mm) into a packaged module (Fig. 4). We mount the chip on a thermo-electric cooler (TEC) and attach a lensed fiber to the input waveguide (with estimated loss below 2 dB). We wirebond the heaters to a fanout board and use a high-speed probe to measure the photodiodes.

## Experimental Results

We first investigate the performance of the ring resonators. The waveguide after the rings includes a tap off to measure the spectral response of the rings and infer that they provide the proper dispersion performance. From here, we observe the resonances of each ring on an optical spectrum analyzer (OSA) and use the integrated heaters to tune them (Fig. 5). We see that the transmission spectrum matches the expected profile from Fig. 3(b). The loss across the ~60 GHz bandwidth is around 1 dB.

Next, we demonstrate recovery of dual-polarization, complex-valued signals using the integrated receiver. The experimental setup is shown in Fig. 6. First, an external-cavity laser near 1530 nm is sent to a DP in-phase and quadrature Mach-Zehnder modulator (DP-IQ-MZM) that is driven by a four-channel digital-to-analog converter (DAC) using a Nyquist-shaped 30-GBd QPSK signal of length $2^{12}$. The total launch power before transmission is -2.5 dBm. After propagating through 80 km of SSMF, the signal is amplified and filtered, with an optical signal-to-noise ratio (OSNR) of 27.7 dB and 7.3 dBm of optical power to the integrated receiver. The on-chip photodiodes are biased to -2 V and are measured with a high-speed probe. The electrical signals are then captured at 80 GS/s using two four-channel digital storage oscilloscopes. In the DSP, the state of polarization is determined by estimating the coupling and phase offset parameters of the Jones matrix by minimizing the mean error between the measured and expected intensities of a training sequence. To perform the phase retrieval, we use the GS algorithm [8] with a block length of 4,096 samples at two samples per symbol. The algorithm includes intensity pre-equalization and selective phase reset, as well as amplitude, bandwidth, and pilot symbol constraints to improve convergence.

The measured QPSK constellations for both polarizations are shown in Fig. 7(a-b), with a bit error rate (BER) of $8.8 \times 10^{-4}$ at 27.7 dB OSNR, which is below the 7% forward error correction (FEC) threshold. Additionally, we transmit DP-8QAM signals at 20 GBd and apply the phase retrieval algorithm. We measure 8QAM constellations for both polarizations (Fig. 7(c-d)) with BER $5.1 \times 10^{-3}$ at 31.1 dB OSNR. The BERs for both modulation formats versus OSNR are plotted in Fig. 7(e). We believe the photodiode probing setup presently limits the bandwidth and could be improved to lower the BER further, but nevertheless the phase retrieval is robust and achieves BER below the FEC threshold.

**Conclusion**

We have presented the first demonstration of a silicon photonic phase retrieval receiver which recovers the full field without a carrier or LO. The dispersion and delay lines needed for the space-time diversity phase retrieval algorithm are implemented by carefully designing the silicon waveguide components. We show transmission of high-speed DP-QPSK and DP-8QAM signals over 80 km of fiber with low BERs. Because it does not require a carrier, our approach is compatible with conventional coherent transmitters. Compared to coherent systems, direct detection-based phase retrieval could allow for a colorless transmitter and therefore an uncooled laser, reducing power consumption [9]. The low cost and small footprint of such a receiver in silicon photonics highlights our approach's benefits in optical networks from data centers to mobile fronthaul.

## Acknowledgements

We wish to thank Di Che, Xi Chen, Nicolas K. Fontaine, and Robert Borkowski for fruitful discussions.

**Figures**

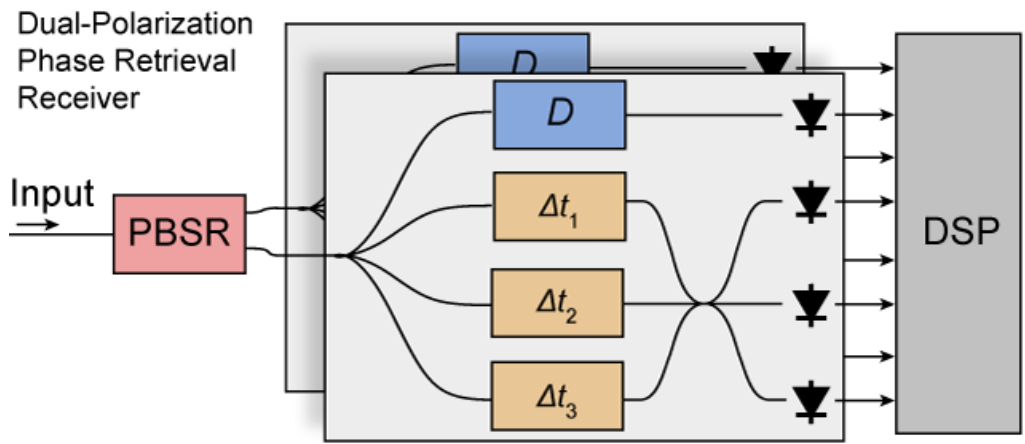

**Figure 1.** Schematic of a phase-retrieval receiver, including elements providing dispersion $D$, time delays $\Delta t$, photodiodes, and digital signal processing (DSP).

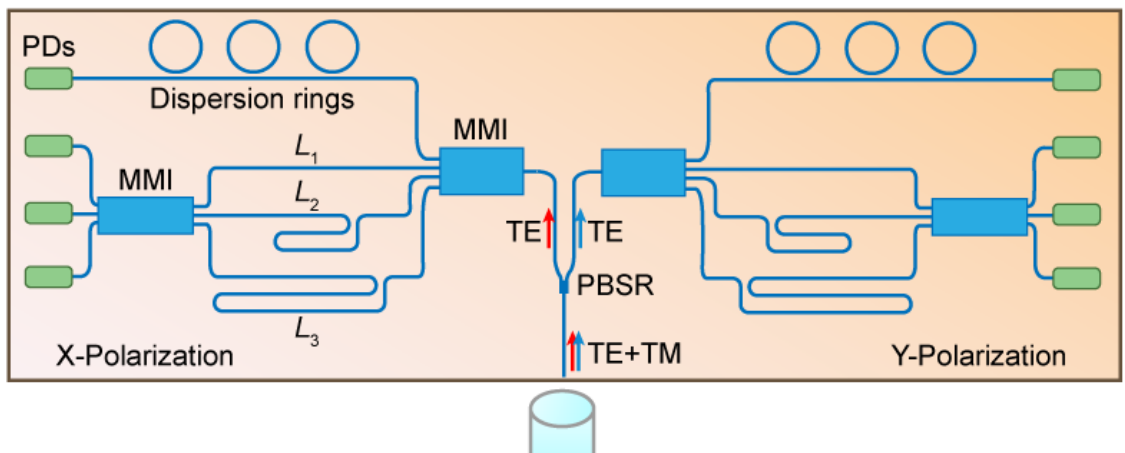

**Figure 2.** Diagram of the integrated phase retrieval receiver.

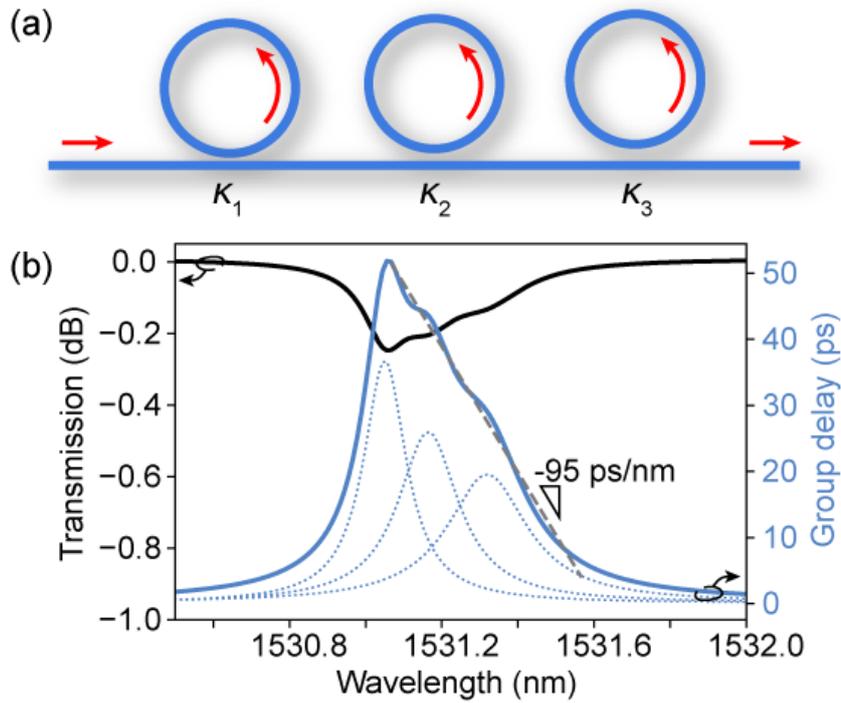

**Figure 3.** (a) Diagram of a set of microrings along a bus waveguide, (b) Simulation of the transmission and group delay of the set of rings. The group delay contributions of the individual rings are shown as dotted lines.

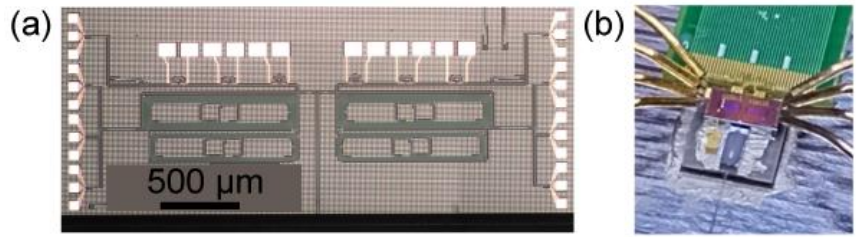

**Figure 4.** (a) Microscope image of the fabricated receiver chip. (b) Photograph of the assembled module.

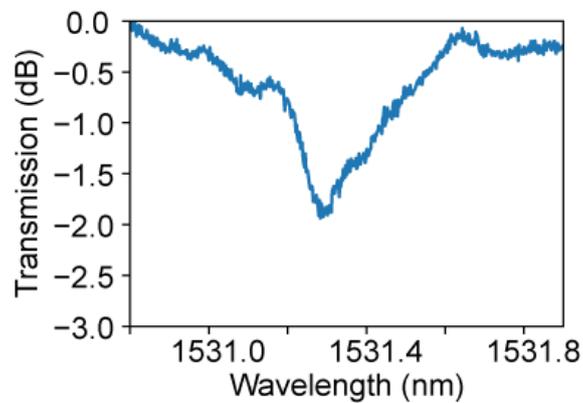

**Figure 5.** OSA measurement of the rings' spectrum.

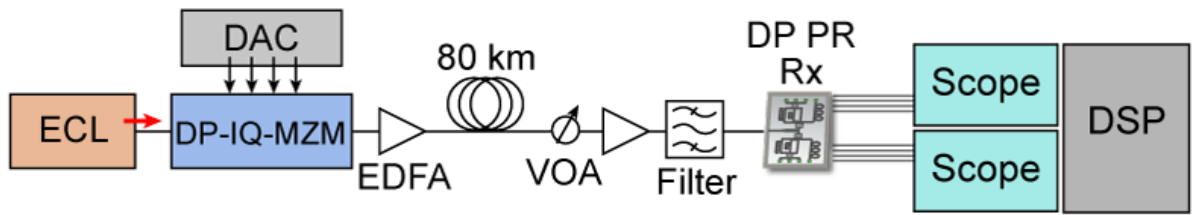

**Figure 6.** Diagram of the experimental setup for the phase retrieval receiver. ECL: external cavity laser, EDFA: erbium doped fiber amplifier, VOA: variable optical attenuator.

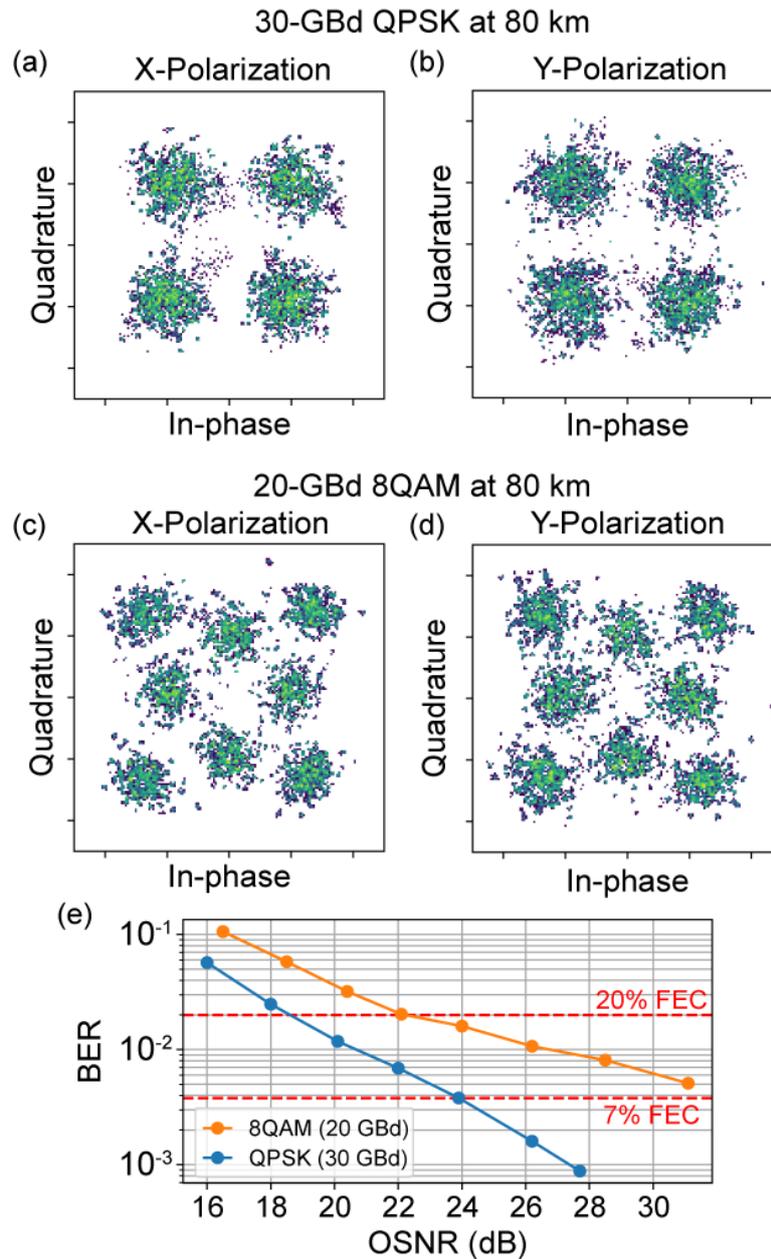

**Figure 7.** (a-d) Measured QPSK and 8QAM constellations for each polarization, (e) Measured BER vs. OSNR.